\documentstyle[prl,aps,psfig]{revtex}

\begin{document}
\draft

\twocolumn[\hsize\textwidth\columnwidth\hsize\csname@twocolumnfalse\endcsname

\title{Structure and Dynamics of Liquid Iron under \\
Earth's Core Conditions}

\author{D. Alf\`{e}$^1$, G. Kresse$^2$ and M. J. Gillan$^3$}
\address{
$^1$Research School of Geological and Geophysical Sciences \\
Birkbeck College and University College London \\
Gower Street, London WC1E 6BT, UK \\
$^2$Institut f\"ur Materialphysik, Universit\"at Wien,
Strudlhofgasse 4, A-1090 Wien, Austria\\
$^3$Physics and Astronomy Department, University College London \\
Gower Street, London WC1E 6BT, UK
}

\maketitle

\begin{abstract}
First-principles molecular dynamics simulations
based on density-functional theory
and the projector augmented wave (PAW) technique have been
used to study the structural and dynamical properties of liquid
iron under Earth's core conditions. As evidence for the accuracy
of the techniques, we present PAW results for a range
of solid-state properties of low- and high-pressure iron,
and compare them with experimental values and the results
of other first-principles calculations. In the liquid-state simulations,
we address particular effort to the study of finite-size effects,
Brillouin-zone sampling and other sources of
technical error. Results
for the radial distribution function, the diffusion
coefficient and the shear viscosity are presented for a
wide range of thermodynamic states relevant to
the Earth's core. Throughout this range, liquid iron is a close-packed
simple liquid with a diffusion coefficient and viscosity
similar to those of typical simple liquids under ambient
conditions.
\end{abstract}
]

\section{Introduction}
The aim of this work is to use first-principles simulation
to determine some of the basic properties of liquid iron at
temperatures and pressures relevant to the Earth's core, and
to present tests and comparisons which establish the reliability and
robustness of the techniques employed. Many lines of evidence
show that the Earth's core consists mainly of iron, with a minor
fraction of light impurities~\cite{birch52,anderson89,poirier91}.
Since the inner part of the core is
solid and the outer part is liquid, a firm grip on the properties
of both solid and liquid iron
under the relevant conditions is crucial to understanding the
behaviour of the core, including its convective dynamics,
heat transport, and the generation of the Earth's magnetic
field.

The laboratory investigation of iron under core conditions is
exceedingly difficult, because of the high pressures and temperatures
needed. At the boundary between the mantle and the core, at a depth of
$\sim$~3000~km, the pressure is 135~GPa and the core temperature is
believed to be $\sim 4000$~K~\cite{anderson97}, while at the boundary
between inner and outer core the pressure is 330~GPa and the
temperature is thought to be in the region of 5000~K. Experiments
using the diamond anvil
cell~\cite{williams87,boehler90,boehler93,andrault97,shen98} can be
performed up to $\sim 200$~GPa, and x-ray diffraction measurements on
solid Fe have been made at conditions approaching the core-mantle
boundary~\cite{andrault97,shen98}.  Beyond this, shock measurements
have given some information about the thermodynamic properties of
solid and liquid iron~\cite{jeanloz79,brown86,bass87}, but the
temperature calibration of these measurements is difficult, and there
are substantial disagreements between the results. The shear viscosity
of the liquid is important for understanding core dynamics, but has
been extremely controversial, with estimates from different sources
spanning many orders of magnitude~\cite{secco95,brazhkin98}. The
laboratory measurement of viscosities using the diamond anvil cell is
becoming feasible up to pressures of $\sim
10$~GPa~\cite{dobson96,leblanc96}, but great technical problems still
need to be overcome before such measurements can be done under core
conditions.

First-principles calculations based on density
functional theory (DFT)~\cite{car85,jones89,gillan97}
are becoming increasingly important in the study of materials
under extreme conditions, and a substantial effort has already
been devoted to iron. It is well established that DFT accurately
reproduces the properties of low-temperature body-centred cubic (b.c.c.)
iron at ambient pressures, including the equilibrium lattice
parameter, bulk modulus and magnetic
moment~\cite{stixrude94,vocadlo97}, and phonon frequencies~\cite{vocadlo99}.
There has also been much DFT work on different crystal
structures of Fe at high pressures, and experimental
low-temperature results for the pressure as a function of
volume $p(V)$ up to $p = 300$~GPa for the hexagonal close-packed
(h.c.p.) structure are accurately
predicted~\cite{stixrude94,vocadlo97,soderlind96}. Further evidence for
the accuracy of DFT comes from the successful prediction
of the transition pressure from the b.c.c. to the h.c.p.
phase~\cite{stixrude94,soderlind96}. Building
on these successes, there have been extensive
investigations~\cite{vocadlo99,soderlind96}
of the relative stability and properties of the main
candidate crystal structures at pressures up to core values.

Very recently, we have applied the pseudopotential/plane-wave
formulation of DFT to calculate the properties of both solid and liquid
iron under core
conditions~\cite{vocadlo97,dewijs98,alfe98a,alfe99}.
The great advantage of this DFT formulation is
that systems containing tens or hundreds of atoms can be treated.
Furthermore, the forces on the atoms are readily calculated, so that
dynamical simulations can be performed, and liquids in thermal
equilibrium can be simulated. We have reported a brief study of
the liquid structure~\cite{vocadlo97}, 
which showed that it is close packed, with a
coordination number slightly above 12. We also reported values
for the diffusion coefficient, and we used the approximate
Stokes-Einstein relation to infer values of the
viscosity~\cite{dewijs98}. Our
results indicated that the viscosity of {\em l}-Fe under core
conditions is roughly 10 times greater than that of typical
simple liquid metals under ambient conditions. However, our liquid Fe
simulations were performed on a very small system of 64 atoms,
and we were unable at that stage to investigate system-size errors
or errors due to other technical factors such as $k$-point
sampling. In addition, we studied only a very small number of
thermodynamic states.

Here we report the results of much more extensive first-principles
simulations covering a wide range of thermodynamic conditions,
and we also describe the thorough study
of technical issues that we have now made.
An important advance over our previous work is that most
of the present calculations are based on the projector
augmented wave (PAW)
formulation of DFT~\cite{blochl94,kresse99}. This is an all-electron
technique which is closely related both to other standard all-electron
techniques like the linear augmented-plane-wave
(LAPW) method~\cite{wei85},
and also to the ultrasoft pseudopotential method used in our
previous work~\cite{vocadlo97,dewijs98,alfe98a,alfe99}. However,
in contrast to other all-electron
methods, PAW allows one to do dynamical first-principles
simulations of the kind that are routinely done with
the pseudopotential approach. We report detailed tests on different
crystal structures of Fe, which show that the PAW
technique reproduces very accurately the available experimental
data, as well as the results of previous all-electron calculations.
We describe the results of our liquid-state PAW simulations on a range of
different system sizes, which allow us to give a quantitative assessment
of size errors and other technical factors.
A further advance over our earlier work is that
we now calculate the viscosity directly from the Green-Kubo relation
involving the stress autocorrelation
function~\cite{alfe98b}, rather than
estimating it from the diffusion coefficient.

Details of our DFT techniques are given in the next Section, and
the results of our tests on crystalline iron are reported
in Sec.~3. Our detailed tests on the reliability
of our techniques for studying the liquid are presented
in Sec.~4, where we also report results
for the radial distribution function, diffusion coefficient
and viscosity of the liquid over a range of conditions. Discussion
and conclusions are given in Sec.~5.

\section{Methods}\label{sec:methods}
Density-functional theory is a general and extremely widely
used set of techniques for treating the energetics of condensed
matter, and it has frequently been reviewed (see e.g.
Refs.~\cite{jones89,gillan97,parr89,payne92}). Originally
formulated as a method for calculating the energy
of the electronic ground state of collections
of atoms for given nuclear positions~\cite{hohenberg64,kohn65},
it was later generalized to calculate the electronic
free energy when the electrons are in thermal equilibrium
at some finite temperature \cite{mermin65} 
(again, for given nuclear positions).
Since DFT also yields the forces on the nuclei {\em via} the
Hellmann-Feynman theorem, it is also possible to perform
first-principles molecular dynamics~\cite{car85}, 
in which the nuclear positions
evolve according to classical mechanics under the action
of the forces, while the electronic subsystem follows adiabatically,
either in the ground state or in the state of instantaneous thermal
equilibrium.

Implementations of DFT can be divided into all-electron
(AE) methods, in which both core and valence electrons are treated
explicitly, and pseudopotential methods, in which only valence
electrons are treated explicitly, their interactions with the cores
being represented by a pseudopotential. AE methods, such as the
well-known FLAPW (full-potential linear augmented plane-wave) and
FP-LMTO (full-potential linear muffin-tin orbitals) schemes, are
traditionally regarded as more `rigorous', but at present can be
applied only to rather small numbers of atoms, and it is difficult to
use them for dynamical simulations. The pseudopotential approach, by
contrast, is routinely used for dynamical simulations on systems of
tens or hundreds of atoms. First-row and transition-metal elements
used to be far more demanding with the pseudopotential approach, but
the introduction of so-called `ultrasoft' pseudopotentials (USPP)
\cite{vanderbilt90} has largely overcome these problems.

In the last few years, a new method, known as PAW (projector
augmented wave) has been developed~\cite{blochl94}, which effectively bridges
the divide between AE and pseudopotential methods. It is an
all-electron method, in the sense that it works with the true Kohn-Sham
orbitals, rather than orbitals that are `pseudized' in the core regions,
and it has the same level of rigor as AE methods such as FLAPW.
At the same time, it is very closely related to the USPP
technique, and reduces to this if certain well-defined
approximations are made, as shown in the recent
analysis by Kresse and Joubert~\cite{kresse99}. The nuclear forces
can be calculated in the same way as in pseudopotential methods,
so that dynamical simulations can be performed at essentially
the same cost as with the USPP method. An extensive
set of comparisons between the PAW, USPP and AE approaches
has recently been presented~\cite{kresse99} for a variety
of small molecules and bulk crystals, including Fe, Co and Ni, and it is
was shown that the three approaches give virtually identical
results in most cases. The only significant exception is
ferromagnetic Fe, where fully converged results are easier to
obtain with PAW than with USPP.
The present work makes direct use of the work of Kresse and Joubert,
and was performed with the PAW facility developed in the VASP
code~\cite{kresse96a,kresse96b}. The details of the core radii,
augmentation-charge cut-offs etc. in the present PAW work are exactly
as in the Fe calculations reported in Ref.~\cite{kresse99}.

It is known that under Earth's core pressures it is not accurate
enough to neglect the response of the 3$s$ and 3$p$ electrons to the
high compression. The 3$p$ response is most significant, and in order
to achieve good accuracy we treat both 3$p$ and 3$d$ as valence
electrons.  However, this makes it very costly to do long dynamical
simulations on large systems, and we showed
earlier~\cite{vocadlo97} that it is an accurate
approximation to treat the 3$p$ orbitals as rigid provided the effect
of their response is included by an effective pair potential. (We
stress that non-linear core corrections are included everywhere in the
present work.) The procedure used to construct this effective
potential is as follows (see also Ref.~\cite{ballone89}).
We calculate the total energies of an
perfect h.c.p. Fe crystal in two ways: in the first we include
the 3$p$ orbitals in valence, and in the second we freeze them in the
core. The effective pair potential is then constructed so as to reproduce
the difference of the energies between the two sets of calculations.
The accuracy of this procedure for treating 3$p$ response is
demonstrated in Sec.~3 in our comparison of the phonon frequencies of
the f.c.c. phase at high pressure. In the following, we refer to
calculations in which 3$p$ and 3$s$ orbitals and below are in the core
as `Ar-core' calculations, and those in which 3$p$ orbitals are
included in the valence set but 3$s$ orbitals and below are in the
core as `Ne3s$^2$-core' calculations.

In the temperature range $T > 3000$~K of interest here,
thermal excitation of electrons is important, and throughout this
work we use the finite-temperature formulation of DFT~\cite{mermin65}.
This means that each Kohn-Sham orbital has an occupation
number given by the usual Fermi-Dirac formula. This thermal
excitation has an important effect both on the pressure in the
system and on the nuclear forces, as will be reported in
detail elsewhere. However, it should be noted that this
procedure ignores the possible temperature dependence of the
exchange-correlation functional about which little is currently
known.

Further technical details are as follows. All the calculations
presented here are based on the form of generalized gradient
approximation (GGA) known as Perdew-Wang
1991~\cite{wang91,perdew92}. In the very few spin-polarized
calculations we have performed, the spin interpolation of the
correlation energy due to Vosko {\em et al.} was
used~\cite{vosko80}. Brillouin-zone sampling was performed using
Monkhorst-Pack (MP) special points~\cite{monkhorst76}. Most of the
dynamical simulations on the liquid were performed using
$\Gamma$-point sampling only, though we shall report tests using more
$k$-points. The plane wave cut-off was 300 eV in all calculations. The
extrapolation of the charge density from one step to the next was
performed using the technique described by Alf\`e~\cite{alfe99b}.  The
time-step used in the dynamical simulations was 1 fs.

\section{Solid iron}
In order to demonstrate the robustness and reliability of the present
PAW methods, we have calculated a range of solid-state properties of
iron and compared them with experimental values, and with the results
of other DFT implementations, including FLAPW, FP-LMTO and ultra-soft
pseudopotentials. We also probe the effect of treating the
valence-core split in different ways.  We have taken pains to ensure
that every result given by the present calculations is fully converged
with respect to $k$-point sampling, plane-wave cut-off, and all other
technical factors.

In Table~\ref{tab:solid}, we report values of the equilibrium volume
$V_0$ for the b.c.c. and h.c.p. structures, the bulk modulus $K_0$ and
its pressure derivative $d K_0 / d p$, and the magnetic moment $\mu$
per atom of the (ferromagnetic) b.c.c. phase. In the calculation of
$K_0$ and $K_0^\prime$ for b.c.c., the moment $\mu$ is, of course,
free to change with volume. In the h.c.p. case, the equilibrium value
of the $c/a$ ratio is determined by minimizing the total energy at
fixed volume. Since in principle $c/a$ can depend on volume, this must
be taken into account in the calculations of $K_0$ and
$K_0^\prime$. In the region of zero pressure we find the value $c/a$ =
1.58, which agrees exactly with the calculated value of
Ref.~\cite{stixrude94}, and is in reasonable agreement with the
experimental values in the range $1.58 - 1.61$ reported in
Ref.~\cite{mao90}.We note the excellent stability of the results with
respect to DFT implementation, and the generally very good agreement
with experimental values. We expect the most accurate variant of our
PAW calculations to be the one based on the Ne3$s^2$ core, and $V_0$
values calculated in this way agree with FLAPW values to within 1~\%
or better. The agreement between PAW[Ne]3$s^2$ and FLAPW results for
$K_0$ and $K_0^\prime$ is also excellent for h.c.p. .  The magnetic
moment depends only weakly on the DFT method. The agreement of our PAW
results with experimental values is very good for b.c.c., but less
good for h.c.p.. However, the experimental values for $V_0$ and $K_0$
in the h.c.p. phase do not come from direct measurements at zero
pressure (since the h.c.p. crystal is unstable at this pressure), but
from an extrapolation from measurements at higher pressures. The
reliability of the ``experimental'' values is therefore not clear.

\begin{figure}
\centerline{\psfig{figure=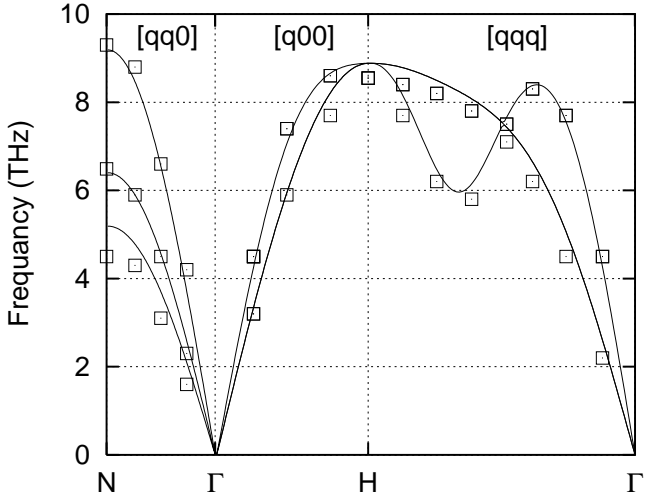,height=2.6in}}
\vskip 10pt
\caption{Phonon dispersion curves of ferromagnetic b.c.c. Fe at zero
pressure along the [100], [110] and [111] directions. Curves show calculated
phonon frequencies, open squares are experimental data of 
Ref.~\protect\cite{brockhouse67}. }\label{fig:phon}
\end{figure}

Our calculated phonon dispersion curves for the zero-pressure
b.c.c. phase are compared with experimental frequencies in
Fig.~\ref{fig:phon}. The calculations were done using our
implementation of the small-displacement method described in
Ref.~\cite{kresse95}.  Small systematic differences can be seen for
transverse modes at some wavevectors on the $\Gamma - N$ and $\Gamma -
H$ lines, but even at worst these do not exceed {\em ca.}~10~\%.

\begin{figure}
\centerline{\psfig{figure=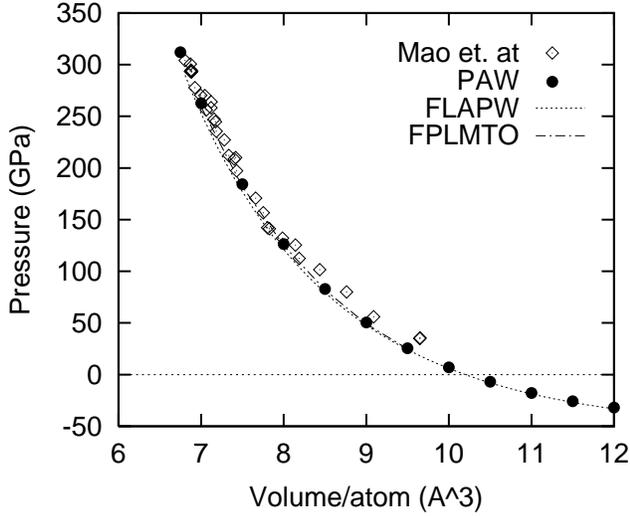,height=2.6in}}
\vskip 10pt
\caption{Pressure as a function of atomic volume of h.c.p. Fe. Solid circles
are present PAW calculations; dotted and chain curves are FLAPW and FP-LMTO
results of Ref.~\protect\cite{stixrude94} and~\protect\cite{soderlind96} 
respectively, diamonds are experimental values of 
Ref.~\protect\cite{mao90}}\label{fig:p_of_v}
\end{figure}

Fig.~\ref{fig:p_of_v} compares calculated values for $p$ as a function
of $V$ of the h.c.p. crystal with experimental measurements due to Mao
{\em et al.}~\cite{mao90}.  On the scale of the Figure, the PAW
results obtained with the Ne3$s^2$ core are indistinguishable from those
given by the Ar core plus pair potential, and are also virtually
identical to the FP-LMTO results of S\"{o}derlind {\em et
al.}~\cite{soderlind96}. There is a detectable difference from the
FLAPW results of Stixrude {\em et al.}~\cite{stixrude94}, but this is
smaller than the difference from the experimental values, which itself
is very small in the pressure range $100 - 300$~GPa of main interest in
this paper.

A further test concerns the low-temperature coexistence pressure of
the b.c.c. and h.c.p. phases, which experimentally is in the range $10
- 15$~GPa~\cite{jephcoat86}. The earlier FLAPW and FP-LMTO
calculations both gave transition pressures of {\em ca.}~11~GPa. The
present PAW calculations give 10, 12 and 13~GPa for the PAW[Ne]3$s^2$,
PAW[Ar] and PAW[Ar] + effective pair potential respectively, in
reasonable agreement with other values.

Finally, we report our comparison of phonon frequencies of the
high-pressure non-magnetic f.c.c. phase with the FP-LMTO calculations
of S\"{o}derlind {\em et al.}~\cite{soderlind96}. Frequencies were
computed at the three volumes and four wavevectors studied in the
FP-LMTO work, and their values are compared in
Table~\ref{tab:fccphon}. We note that this is not a very direct
comparison, since the FP-LMTO calculations were based on the LDA
rather than the GGA used here. We have shown
elsewhere~\cite{vocadlo97} that the frequencies of f.c.c. Fe increase
slightly when LDA is replaced by GGA, and this is consistent with the
difference shown in the Table. However, the differences are no more
than a few percent.  We also note that the PAW frequencies obtained
with the Ar core plus pair potential are almost identical to those
given by the Ne3$s^2$ core, the effect of the pair potential is to
increase the frequencies by 5-10\%. This is useful evidence that the
procedure of using the Ar core plus pair potential gives a good
account of vibrational as well as equilibrium properties.

Our overall conclusion from these tests is that our present PAW
calculations appear to reproduce very accurately the properties
of real crystalline Fe, and the differences from pseudopotential
and standard all-electron implementations of DFT are
extremely small. This provides a firm foundation for our
PAW simulations of liquid Fe presented in the next Section.

\section{Liquid iron}
\subsection{A representative ab initio simulation}
\label{sec:rep}
For orientation purposes, we present first the results obtained
from a direct {\em ab initio} simulation of liquid
iron at the temperature $T = 4300$~K and the density
$\rho = 10700$~kg~m$^{-3}$, which corresponds roughly
to conditions at the core-mantle boundary. This simulation
was performed on a system of 67 atoms, and used $\Gamma$-point
sampling. The system was initiated and thoroughly equilibrated
in a way that will be described below. The duration of the
simulation after equilibration was 15~ps. The pressure
was calculated in the course of the simulation and its
value was 132~GPa, which should be compared with the
value 135~GPa at the core-mantle boundary.

\begin{figure}
\centerline{\psfig{figure=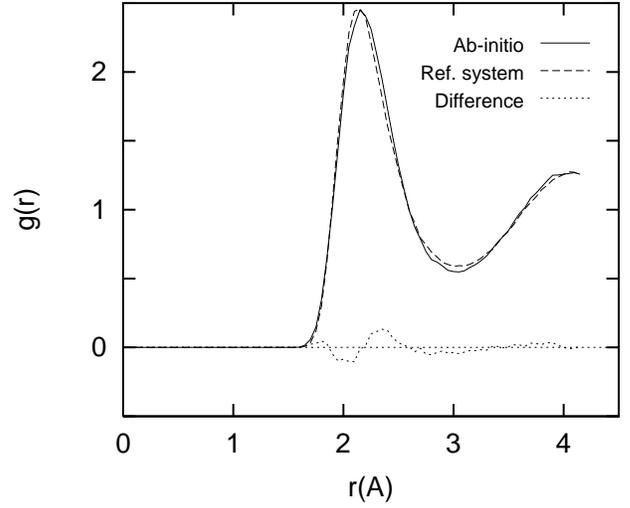,height=2.6in}}
\vskip 10pt
\caption{Radial distribution function $g(r)$ of {\em l}-Fe
at $T = 4300$~K and
$\rho = 10700$~kg~m$^{-3}$. Solid and dashed curves are {\em ab-initio} and
reference-model results, dotted curve is the difference of the two.}
\label{fig:gr}
\end{figure}

The radial distribution function $g(r)$ from this simulation
is displayed in Fig.~\ref{fig:gr}. It shows the form
typical of real and theoretical simple liquids such as liquid
aluminum or the Lennard-Jones liquid. It is instructive to
calculate the average coordination number $N_{\rm c}$ characterising
the number of nearest neighbours surrounding each atom. This is
defined as:
\begin{equation}
N_{\rm c} = 4 \pi n \int_0^{r_{\rm c}} dr \, r^2 g (r)  \; ,
\end{equation}
where $n$ is the atomic number density and
$r_{\rm c}$ is the position of the first minimum in
$g(r)$. We find the value $N_{\rm c} = 13.2$.
This is actually
slightly greater than the coordination number of 12 exhibited by the
close-packed h.c.p. and f.c.c. structures, and it is clear that the liquid
structure displays a very dense form of packing. We note
that by no means all monatomic liquids show this high coordination.
Directionally bonded liquids such as {\em l}-Se and {\em l}-Si
have very much lower coordination numbers ($N_{\rm c} \simeq 2$
and 6 in these two cases). The high value of $N_{\rm c}$ we find
for high-pressure {\em l}-Fe is evidence of strong repulsive
forces between the atoms and a complete absence of directional
bonding. This is, of course, not surprising given the
very high degree of compression.

\begin{figure}
\centerline{\psfig{figure=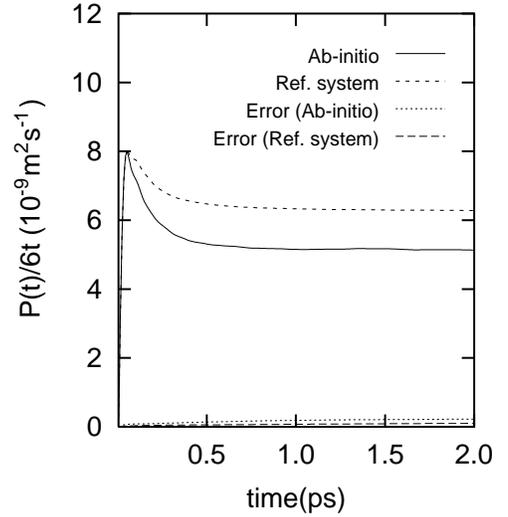,height=2.6in}}
\vskip 10pt
\caption{Mean-square displacement $P(t)$ divided
by $6t$ for {\em l}-Fe at $T = 4300$~K
and $\rho = 10700$~kg~m$^{-3}$. Solid and short-dashed curves show results for
{\em ab-initio} and reference system, with
statistical error shown as dots and long dashes.}\label{fig:msq}
\end{figure}

We now pass to the self-diffusion coefficient $D$, which we obtain
from the asymptotic slope of the time-dependent mean-square
displacement (MSD) $\langle \mid {\bf r}_i (t + t_0 ) -
{\bf r}_i ( t_0 ) \mid^2 \rangle$:
\begin{equation}
P(t) = \langle \mid {\bf r}_i ( t + t_0 ) -
{\bf r}_i ( t_0 ) \mid^2 \rangle
\rightarrow
6 D \mid t \mid 
\end{equation}
in the long-time limit $\mid t \mid \rightarrow \infty$.  The quantity
$P(t)/6t$ from our simulation is reported in Fig.~\ref{fig:msq}. We also
report in the Figure the statistical errors on the MSD, which we
shall need to refer to later; these errors were estimated by
calculating the MSD separately for every atom in the system and
examining the scatter of the results.  As generally happens in simple
liquids, the asymptotic linear region in the MSD is attained extremely
quickly, after a transient period of only $\sim 0.3$~ps. The
asymptotic slope gives the value $D = (5.2 \pm 0.2) \times
10^{-9}$~m$^2$~s$^{-1}$.  The value of $D$ is quite typical of simple
liquids under ambient conditions. For example, the diffusion
coefficient of liquid Al near its triple point~\cite{lide93} is
$\sim 8 \times 10^{-9}$~m$^2$s$^{-1}$. This indicates that the
dynamics of the atoms in {\em l}-Fe under the present thermodynamic
conditions is not drastically affected by the high degree of
compression.

The final quantity we examine is the shear viscosity $\eta$. In our
earlier work~\cite{dewijs98}, we relied on the Stokes-Einstein relation
to give a rough estimate of the viscosity, but we
subsequently showed that this indirect method is completely
unnecessary, since the direct, rigorous calculation of
$\eta$ using the Green-Kubo relation is perfectly feasible~\cite{alfe98b}.
According to this relation, $\eta$ is given by:
\begin{equation}\label{eqn:sacf}
\eta = \frac{V}{k_{\rm B} T} \int_0^\infty dt \,
\langle P_{x y} (t) P_{x y} (0) \rangle \; ,
\end{equation}
where $\langle P_{x y} (t) P_{x y} (0) \rangle$ is the
stress autocorrelation function (SACF), i.e.
the autocorrelation function of the off-diagonal
component $P_{x y}$ of the stress tensor
at times separated by $t$, and $V$ is the volume
of the system. Techniques for calculating the SACF are presented
in Ref.~\cite{alfe98b}, where it is noted that statistical accuracy
is improved by taking the average $\phi (t)$
of the five independent correlation functions of the traceless stress
tensor $P_{x y}$, $P_{y z}$, $P_{z x}$,
$\frac{1}{2} ( P_{x x} - P_{y y} )$ and
$\frac{1}{2} ( P_{y y} - P_{z z} )$.

The average SACF $\phi (t)$ and its time-integral
$\int_0^t d t^\prime \,
\phi ( t^\prime )$ for the present thermodynamic
state are reported in Fig.~\ref{fig:sacf}, with error bars for the
time integral. We obtain from this the value
$\eta = 8.5 \pm 1$~mPa~s. It is interesting
to compare this with the value that
would be obtained from the Stokes-Einstein relation:
\begin{equation}
D \eta = k_{\rm B} T / 2 \pi a \, ,
\end{equation}
where $a$ is an effective atomic diameter.  In our previous work, we
took the effective atomic diameter $a$ to be the nearest-neighbour
distance in the close-packed solid having the same density, which in
the present case is 2.15~\AA. Using the value $D = 5.2 \times
10^{-9}$~m$^2$s$^{-1}$ reported above, we obtain the estimate $\eta =
8.5$~mPa~s, which happens to be the same as the Green-Kubo value.

Useful though they are, the results we have just presented are
limited in two ways. First, there are technical limitations.
Our results have been obtained from a rather small simulated
system of only 67 atoms, and we must clearly
try to show that they are not seriously influenced by size
effects. Similarly, we need to know that the results are not
affected by other technical factors, such as the use of
$\Gamma$-point sampling, the choice of the PAW technique rather
than some other {\em ab initio} technique, or the split that we
have chosen between valence and core states. The second limitation
is that we have studied only one thermodynamic state. We want
to know how the properties of {\em l}-Fe vary over the range of
conditions relevant to the Earth's core. In overcoming both
these kinds of limitation, the notion of a reference
system will be very important, and we discuss this next.

\subsection{The reference system}
Since {\em ab initio} simulation is very costly, it is difficult
to study size effects by directly simulating large systems.
This is particularly difficult for dynamical properties like
the diffusion coefficient and the viscosity, since rather long
simulations are needed. It will be therefore very helpful to
have available a simple model or reference system which closely mimics
the behaviour of the full {\em ab initio} system. We have in mind
a model in which the atoms interact through a simple
pairwise potential $\phi (r)$, so
that its total potential energy $U_0$ is
given by:
\begin{equation}
U_0 = \frac{1}{2} \sum_{i \ne j} \phi ( \mid {\bf r}_i
- {\bf r}_j \mid ) \; .
\end{equation}
If the reference system is to behave like the {\em ab initio}
system, then its total energy $U_0$ must closely resemble
the {\em ab initio} total energy $U$. This means that the
difference $\Delta U \equiv U - U_0$ must be small. In fact, the
requirement is slightly weaker than this. It makes no difference
to the thermal-equilibrium structure or dynamics of the liquid
if we add a constant to the total energy. So the requirement
is really that the {\em fluctuations} of
$\Delta U$ should be small. We do not need to demand that the strength
of these fluctuations be small over the whole of configuration
space. It suffices that they are small over the region explored
by the atoms. Regions of configuration space rarely visited
by the system should be given a low weighting in assessing
the fluctuations, and the natural way to do this is to weight
configurations with the Boltzmann factor
$\exp [ - \beta U ( {\bf r}_1 , \ldots {\bf r}_N ) ]$. The
requirement is therefore that the thermal average of the mean-square
fluctuations of $\Delta U$ defined by $\langle ( \Delta U -
\langle \Delta U \rangle )^2 \rangle$ be as small as possible.
Here, the thermal average $\langle \, \cdot \, \rangle$ can be
evaluated as a time average over  configurations generated
in the {\em ab initio} simulation. It will be noted that this
requirement depends on thermodynamic state, so that the best
model system may be different for different states.

We have every reason to expect that our {\em ab initio} liquid can be
well modelled by a simple reference system, since we have seen that
its properties closely resemble those of typical simple
liquids. Initially, we attempted to use the Lennard-Jones model, in
which $\phi (r) = 4 \epsilon ( ( \sigma / r )^{12} - ( r / \sigma )^6
)$, where $\sigma$ and $\epsilon$ characterise the atomic diameter and
the depth of the attractive potential well respectively. We found that
it is possible to choose $\sigma$ and $\epsilon$ so that the
fluctuations $\delta \Delta U \equiv \Delta U - \langle \Delta U
\rangle$ are reasonably small: the minimum rms value that we achieved
was $[ \langle ( \delta \Delta U )^2 \rangle / N ]^{1/2} = 0.18$~eV ($N$
is the total number of atoms in the system). However, the rdf of the
resulting LJ liquid differs appreciably from that of the {\em ab
initio} system. Further experimentation showed that a far better
reference system is provided by a pure inverse-power potential $\phi
(r) = B / r^\alpha$. After adjusting the parameters $B$ and $\alpha$
so as to minimise the strength of the fluctuations $\delta \Delta U$,
we achieved a much improved rms value $[ \langle ( \delta \Delta U )^2
\rangle / N ]^{1/2}$ equal to 0.08~eV.  Bearing in mind that $k_B T =
0.37$~eV at the temperature of interest, this represents a very
satisfactory fit to the {\em ab initio} system. The resulting value of
the exponent $\alpha$ is 5.86, and $B$ is such that for $r = 2$~\AA\
the potential $\phi(r)$ is 1.95~eV.

The inverse-power reference system is so close to the
{\em ab initio} system that its rdf $g_0 (r)$ is almost
indistinguishable from the rdf of the {\em ab initio} system. (Naturally,
in making this comparison, we treat the two systems at exactly the same
thermodynamic state and using the same 67-atom repeating cell.)
To show this, we plot the difference $\Delta g (r) \equiv
g (r) - g_0 (r)$ in Fig.~\ref{fig:gr}. The first peak has almost
exactly the same height in the two cases, and the first minimum
is also virtually identical; the difference of the two rdfs
consists mainly of a slight shift of the first
peak of the reference-system to a smaller radius.

Since the reference system is so good, we should expect
it to reproduce also the dynamics of the {\em ab initio}
system. To test this, we compare in Fig.~\ref{fig:msq} the msd
of the reference system with the {\em ab initio} results, using
again the 67-atom repeating system. The two curves agree
very closely for short times, but there is a significant
difference for $t > 0.1$~ps, which is well outside the
statistical errors. The diffusion coefficients obtained for
PAW and the reference model are $5.2 \times 10^{-9}$ and
$6.1 \times 10^{-9}$~m$^2$s$^{-1}$ respectively. Given
that the peak position of $g(r)$ is at a slightly
smaller radius for the reference model, it is perhaps
surprising that the model makes the atoms more mobile,
but one should remember that the rdf describes only 2-body
correlations, and higher correlations may well be important
for diffusion.

Finally, we make the same comparisons for the stress-stress correlation
function and its time integral (see Fig.~\ref{fig:sacf}). The asymptotic
value of the viscosity integral is significantly
smaller for the reference model -- as one
would expect from the higher diffusion coefficient in this case.
The {\em ab initio} and reference viscosities are 8.5 and 7.0~mPa~s
respectively, so that they differ by {\em ca.}~20~\%, as would be
expected from the Stokes-Einstein relation.

Our overall conclusion from these comparisons is that the simple
inverse-power reference system with appropriately chosen
parameters reproduces the structure of the
{\em ab initio} liquid very well and its dynamics
reasonably well. It will be asked whether the same reference
system also works for other thermodynamic states, or whether
the parameters $B$ and $\alpha$ have to be refitted for
every state. This question will be answered in Sec.~\ref{sec:state}.

\subsection{Size effects}
As an aid to addressing size effects, we have performed
a number of {\em ab initio} simulations at the same thermodynamic
state as before ($T = 4300$~K, $\rho = 10700$~kgm$^{-3}$),
using cell sizes ranging from 89 to 241 atoms. The preparation
and equilibration of these systems were done using the inverse-power
reference system. Since the latter so closely mimics the {\em ab initio}
system for the 67-atom cell, it should provide a well
equilibrated starting point for {\em ab initio} simulation of
larger systems. The duration of all the {\em ab initio}
simulations after equilibration was 1~ps.

\begin{figure}
\centerline{\psfig{figure=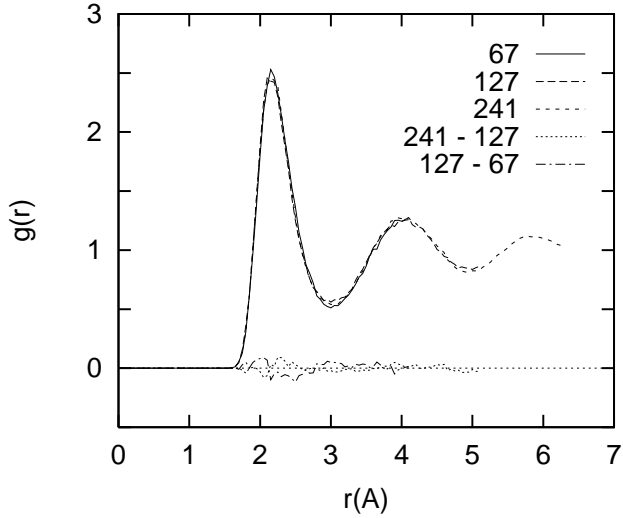,height=2.6in}}
\vskip 10pt
\caption{Comparison of radial distribution function $g(r)$ from
{\em ab-initio} simulations of {\em l}-Fe using
different system sizes. Solid,
long-dash and short-dash curves show $g(r)$ for systems of 67, 127 and 241 
atoms; dotted and chain curves show differences $g_N (r) - g_{N^\prime}$
for $(N,N^\prime)$ equal to (127,67) and (241,127) respectively. }
\label{fig:grsize}
\end{figure}

A duration of 1~ps is enough to give excellent statistical accuracy
for the rdf, so our examination of size effects on $g(r)$ was done
using the {\em ab initio} simulations directly. We find that the
dependence of $g(r)$ on system size is so small that it cannot easily
be seen on simple plots of $g(r)$.  To show these effects, we
therefore plot the differences $g_N (r) - g_{N^\prime} (r)$ between
the rdfs for different numbers of atoms. These differences are
reported in Fig.~\ref{fig:grsize} for $( N , N^\prime )$ equal to $(
127 , 67 )$ and $( 241 , 127 )$.  The difference between $g_{67} (r)$
and $g_{127} (r)$ is small and comes almost entirely in the region of
the first peak; the same is true of the difference between $g_{127}
(r)$ and $g_{241} (r)$.

We have also looked for size effects by studying how well the
inverse-power reference system constructed by fitting to the
67-atom system reproduces the total energy of the larger
systems. We find that the rms strength of the fluctuations
$[ \langle ( \delta \Delta U )^2 \rangle / N ]^{1/2}$
observed in the 1~ps {\em ab initio} simulations for the larger systems
is essentially the same as what we found for the 67-atom system. This
confirms that the reference system mimics the large {\em ab initio}
systems as well as it mimics the small one.

A duration of 1~ps is too short to study size effects on $D$ and
$\eta$ directly from the {\em ab initio} simulations: comparisons
between results for different system sizes would be vitiated by
statistical noise. However, since the reference model appears to work
equally well for all system sizes, we can legitimately use this model
to study size effects. We have therefore calculated $D$ and $\eta$
from simulations of the reference model, using simulations lasting for
500~ps after equilibration. The results are reported in
Table~\ref{tab:finitesize}. We note that $D$ increases slightly as we go to
larger systems. The effect of system size on $D$ in hard-sphere
systems was studied by Erpenbeck and Wood~\cite{erpenbeck85}, who
showed that the calculation of $D$ with 64 atoms would underestimate
its value by ca. 20\%. This is in the same direction as the effect we
are seeing, but in our case the error appear to be less than
10\%. According to the Stoke-Einstein relation, we would expect a
decrease of $\eta$ with increasing system size, but this is not clear
from our results.

\subsection{Other technical factors}

We made tests to assess the influence of Brillouin-zone sampling.
Since {\em ab initio} simulations using many $k$-points are very
demanding, we have adopted an indirect approach. Instead of performing
direct {\em ab initio} simulations with many $k$-points, we have drawn
sets of atomic positions $\{ {\bf r}_i \}$ from the $\Gamma$-point
simulation, and calculated the total energy $U_P$ with $4$ MP 
$k$-points for these configurations. We then compute the rms
strength of the fluctuations $[ \langle ( \delta \Delta U_P )^2
\rangle / N ]^{1/2}$, where $\delta \Delta U_P$ denotes the
fluctuation $U_P - U_1 - \langle U_P - U_1 \rangle$, with $U_1$ the
$\Gamma$-point energy.
This rms fluctuation strength is 0.02 eV, which is much smaller
then the rms fluctuation between the {\em ab-initio} ($\Gamma$-point) system 
and the reference system.
This implies that the expected error in $D$ and $\eta$ would be barely
noticeable, certainly smaller than the statistical errors on these
quantities.

In order to test the reliability of using the Ar core plus effective
pair potential (see Sec.~\ref{sec:methods}), we did a simulation of
the liquid at the same thermodynamic state as before,
but with the Ne3$s^2$ core. The duration of this simulation
was 4.5~ps. The rdf for this simulation was almost identical
to that obtained with the Ar core, the difference being somewhat smaller
than the size-effect difference between $g(r)$ for 67 and 127 atoms
discussed above. Finally, we have compared $g(r)$ obtained in the
present PAW simulations with the one given by our earlier
simulations~\cite{vocadlo97} at the same thermodynamic state, which were
based on ultrasoft pseudopotentials. Here again, the difference is
no bigger than the size effect between 67 and 127 atoms.

Our conclusion from these tests is that our simulations
are completely robust with respect to size effects, $k$-point sampling,
the split between valence and core states, and the use of
PAW rather than the pseudopotential approach.

\subsection{Dependence of liquid properties on thermodynamic state}
\label{sec:state}
In addition to the simulations reported in Sec.~\ref{sec:rep}, we have
also performed PAW simulations at the 15 thermodynamic states listed
in Table~\ref{tab:pressure}, all these simulations being done on the
67-atom system.  With these simulations, we cover the temperature
range $3000 - 8000$~K and the pressure range $60 - 390$~GPa, so that
we more than cover the range of interest for the Earth's liquid
core. The Table reports a comparison of the pressures calculated in
the simulations with the pressures deduced by Anderson and
Ahrens~\cite{anderson94} from a conflation of experimental data.
Our first-principles pressures reproduce the
experimental values to within $2 - 3$~\% at low densities,
but they are systematically too high by {\em ca.}~7~\%
at high densities. It is not clear yet whether the high-density
discrepancies indicate a real deficiency in the {\em ab initio}
calculations rather than problems in the interpretation of
the experimental data. We are currently using free-energy calculations
to study the thermodynamics of the liquid in more detail, and we
hope that this will shed light on the question.

Rather than reporting detailed results for $g(r)$, $D$ and
$\eta$ at each thermodynamic state, we can exploit the properties
of the reference system to present the results in a compact form.
We have taken the reference system fitted to the {\em ab initio}
simulations at $T = 4300$~K and $\rho = 10700$~kg~m$^{-3}$, and
examined the fluctuations $\delta \Delta U$ in the {\em ab initio}
simulations done at the other thermodynamic states. We find that
the strength of the fluctuations is not significantly greater
for these other states than it was for the state for which
the model was fitted. This strongly indicates the same reference
model, with exactly the same parameters, mimics the first-principles
system (almost) equally well at all thermodynamic states.

The invariance of the reference system implies a truly remarkable
simplicity in the variation of the liquid properties with
thermodynamic state. It has long been recognised that the
properties of an an inverse-power system in thermal equilibrium
depend non-trivially only on a single thermodynamic variable,
rather than on temperature $T$ and number density $n$
independently~\cite{hoover71}. This variable is:
\begin{equation}
\zeta = B n^{\alpha / 3} / k_B T \; .
\end{equation}
Any other dimensionless combination of the parameters
$B$, $T$ and $n$ is a function only of $\zeta$.
This means, for example, that the rdf $g(r)$, since it is
dimensionless, can depend only on $\zeta$, provided it is
regarded as a function of the dimensionless distance
$\xi \equiv n^{1/3} r$. To formulate this statement precisely,
let us define the `reduced' rdf $\bar{g} ( \xi )$ as:
\begin{equation}\label{eqn:reduced1}
\bar{g} ( n^{1/3} r ) = g (r) . \; 
\end{equation}
Then $\bar{g} ( \xi )$ must be a function only of $\zeta$.
Similarly, the reduced
diffusion coefficient $\bar{D}$ and viscosity $\bar{\eta}$,
defined by:
\begin{equation}\label{eqn:reduced2}
\bar{D} = ( m / k_B T )^{1/2} n^{1/3} D \; \; , \; \; \; \; \;
\bar{\eta} = ( m k_B T )^{-1/2} n^{-2/3} \eta
\end{equation}
can depend only on $\zeta$.

\begin{figure}
\centerline{\psfig{figure=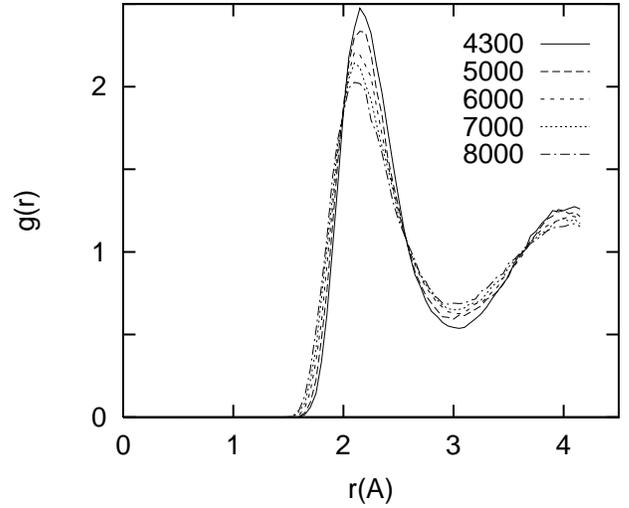,height=2.6in}}
\vskip 10pt
\caption{Variation of $g(r)$ with temperature
from {\em ab-initio} simulations of
{\em l}-Fe at the fixed density $\rho = 10700$~kg~m$^{-3}$. Results are shown
for the five temperatures $T = 4300$, 5000, 6000, 7000 and 8000~K.}
\label{fig:grtemp}
\end{figure}

To show the $\zeta$-dependence of the rdf, we report in
Fig.~\ref{fig:grtemp} the rdf $g(r)$ for the
five temperatures 4300, 5000, 6000, 7000 and 8000~K at the same
density $\rho = 10700$~kg~m$^{-3}$; the corresponding values of
$\zeta$ are 4.51, 3.88, 3.23, 2.77 and 2.43. The effect of varying
temperature (or $\zeta$) is clearly not dramatic, and consists of the
expected weakening and broadening of the structure with increasing
$T$. Since the variation is so regular, we can obtain $\bar{g} ( \xi
)$ for any intermediate value of $\zeta$ by simple linear
interpolation between the curves. With this information, we can then
test explicitly our expectation that $\bar{g} ( \xi )$ depends only on
$\zeta$. To do this, we go to states at other densities $\rho$ and
compare the reduced rdf with the $\bar{g} ( \xi )$ for the same
$\zeta$ but for the density $\rho = 10700$~kg~m$^{-3}$. We find that
the agreement is excellent; indeed, the differences between $\bar{g} (
\xi )$ at the same $\zeta$ but different density are even smaller than
those between $g(r)$ for {\em ab initio} and reference model discussed
in Sec.~\ref{sec:rep}. This shows that the rdf for any thermodynamic
state in the range we have studied can be obtained directly from the
results shown in Fig.~\ref{fig:grtemp}.

Since most of the simulations reported in Table~\ref{tab:d_eta} are rather
short -- typically no more than 4~ps -- the statistical accuracy on
$D$ and $\eta$ is not great. Numerical results, together with error
estimates obtained as in Sec.~\ref{sec:rep}, are reported in
Table~\ref{tab:d_eta}. If the reduced quantities $\bar{D}$ and
$\bar{\eta}$ defined in eqn~(\ref{eqn:reduced2})
depend only on $\zeta$, then a plot of $\bar{D}$ against $\zeta$
should consist of a single curve, and similarly for $\bar{\eta}$.
This hypothesis is tested in Fig.~\ref{fig:d_eta}, and appears to be satisfied
within the (significant) errors. This means that the diffusion
coefficient and viscosity at any thermodynamic state in the range we
have studied can be deduced from the data in Fig.~~\ref{fig:d_eta}.

\section{Discussion and conclusions}
The {\em ab initio} simulations we have presented demonstrate that
under Earth's core conditions {\em l}-Fe is typical simple
liquid. In common with other simple liquids like {\em l}-Ar and
{\em l}-Al, it has a close-packed structure, the coordination number
in the present case being {\em ca.}~13. For the entire pressure-temperature
domain of interest for the Earth's outer core, the diffusion
coefficient $D$ and viscosity $\eta$ are comparable with those
of typical simple liquids, $D$ being
{\em ca.}~$5 \times 10^{-9}$~m$^2$s$^{-1}$ and $\eta$
being in the range $8 - 15$~mPa~s, depending on
the detailed thermodynamic state.

We have shown that the structural properties of {\em l}-Fe are
reproduced very accurately and the dynamical properties fairly
accurately by an inverse-power model, and consequently that they
display a remarkable scaling property. Instead of depending on $T$ and
$n$ separately, their only non-trivial dependence is on the combined
thermodynamic variable $\zeta$ discussed in Sec.~\ref{sec:state}.
Since the Earth's outer core is in a state of convection, the
temperature and density will lie on adiabats. It is straightforward to
show that the variation of $\zeta$ along adiabats will be rather
weak. For example, if we take the data for high-pressure liquid iron
compiled by Anderson and Ahrens~\cite{anderson94}, then the adiabat
for $T = 6000$~K at the ICB pressure 330~GPa has $T = 4300$~K at the
CMB pressure of 135~GPa. Taking the densities at these two points from
the same source, we find values of $\zeta$ equal to 4.85 and 4.51 at
the ICB and CMB respectively. Fig.~~\ref{fig:grtemp} shows that this
small variation of $\zeta$ leaves the liquid structure virtually
unchanged, apart from a trivial length scaling. For all practical
purposes, then, it can be assumed that variation of thermodynamic
conditions across the range found in the core has almost no effect on
the liquid structure. For the same reasons, the reduced diffusion
coefficient $\bar{D}$ and viscosity $\bar{\eta}$ also show little
variation. From the results reported in Sec.~\ref{sec:state}, it is
readily shown that the true (unreduced) diffusion coefficient $D$ is
$5 \pm 0.5 \times 10^{-9}$~m$^2$s$^{-1}$ without significant variation
as one goes from ICB to CMB pressures along the above-mentioned
adiabat, and $\eta$ goes from $15 \pm 5$ mPa~s to $8.5 \pm 1$ mPa~s
along this adiabat.

We have made strenuous efforts to demonstrate the robustness
and reliability of the calculations. Our results on the solid
show that the predictions from our pseudopotential and
PAW {\em ab initio} techniques agree very closely both with
experiment and with the results of other all-electron calculations.
This is true both at ambient pressures and at Earth's core pressures.
For the liquid, we have shown that our results are completely
robust with respect to all the main technical factors: size
of simulated system, $k$-point sampling, choice of {\em ab initio}
method and split between core and valence states.
At present, {\em ab initio} dynamical simulations of the type
presented here are only practicable with the PAW or pseudopotential
techniques. However, our comparisons leave little doubt
that if such simulations were feasible with other DFT
techniques, such as FLAPW, virtually identical results
would be obtained.
In addition,
we have been able to compare the calculated pressure in the
liquid with values deduced from experiments, and again the good
agreement supports the validity of our techniques.

Finally, we note the implications of this work for the controversy
about the viscosity of the Earth's core. In our earlier work on this
problem, we presented {\em ab initio} simulations on a 64-atom
system and used results for the diffusion coefficient together with
the Stokes-Einstein relation to argue that the viscosity of
pure {\em l}-Fe under core conditions is {\em ca.}~13~mPa~s.
Although we regarded this prediction as sound, it was certainly open
to the objection that the simulated system was too small, and that the
viscosity was inferred indirectly. We believe that the present
work has completely overcome both these objections, as well as
giving results for the viscosity over a wide range of conditions.
We fully confirm our earlier conclusion that the viscosity
of pure liquid iron under Earth's core conditions
is at the lower end of the range of proposed values.

In conclusion, {\em l}-Fe under Earth's core conditions is a
simple liquid with a close-packed structure and a coordination
number of 13, and a viscosity going
from 15~mPa~s to 8~mPa~s as conditions
go from the the inner-core boundary to the core-mantle boundary. These
results are completely robust with respect to system-size effects,
choice of {\em ab initio} method, and other technical factors.

\section*{Acknowledgments}
The work of DA was supported by NERC grant GST/O2/1454 to G. D. Price
and M. J. Gillan, and the work of GK by EPSRC grant GR/L08946. Allocations
of time on the Cray T3E machines at the Manchester CSAR service
and at Edinburgh Parallel Computer Centre was provided through the
UK Car-Parrinello consortium and the Minerals Physics consortium.

\clearpage
\begin{table}
\begin{tabular}{llcccc}
 & & $V_0$     & $K_0$ & $ K_0^\prime $ & $\mu / \mu_B$ \\
 & & (\AA$^3$) & (GPa) &         \\
\tableline
b.c.c. & PAW [Ne]$3s^2$     & 11.51 & 178 & 4.8 & 2.22 \\
       & PAW [Ar]           & 11.47 & 176 & 4.8 & 2.23 \\
       & PAW [Ar] + corr    & 11.53 & 176 & 4.8 & \\
       & USPP [Ne]$3s^2$    & 11.63 & 181 & 4.6 & 2.25 \\
       & USPP [Ar]          & 11.74 & 167 & 4.7 & 2.32 \\
       & USPP [Ar] + corr   & 11.78 & 163 & 4.7 & \\
       & FLAPW              & 11.39 & 189 & 4.9 & 2.17 \\
       & expt.              & 11.80 & 162 -- 176 & 5.0 & 2.12 \\

& & & & \\
h.c.p.  & PAW [Ne]$3s^2$   & 10.25 & 287 & 4.5 & \\
        & PAW [Ar]         & 10.19 & 289 & 4.5 & \\
        & PAW [Ar] + corr  & 10.26 & 288 & 4.5 & \\
        & USPP [Ne]$3s^2$  & 10.35 & 291 & 4.4 & \\
        & USPP [Ar]        & 10.32 & 281 & 4.2 & \\
        & USPP [Ar] + corr & 10.38 & 296 & 4.2 & \\
        & FLAPW            & 10.20 & 291 & 4.4 & \\
        & expt.            & 11.2  & 208 &     & \\
\end{tabular}
\caption{Properties of the b.c.c. and h.c.p. crystal structures of Fe at zero 
pressure obtained in the present PAW and USPP calculations compared with
FLAPW and experimental values. Properties reported are: 
the equilibrium volume per atom
$V_0$, the bulk modulus $K_0$, the pressure derivative $K_0^\prime \equiv 
dK_0/dp$ and the magnetic moment $\mu$ per atom (units of Bohr magneton
$\mu_B$). FLAPW results are those of Ref.~\protect\cite{stixrude94}. 
Experimental results all come from Ref.~\protect\cite{knittle95}, except for
$\mu$, which comes from Ref.~\protect\cite{lonzarich80}.
}\label{tab:solid}      
\end{table}

\begin{table}
\begin{tabular}{llcccc}
$V_0$ (\AA$^3$)  &     & L[100] & T[100] & L[111] & T[111] \\
\tableline
6.17 & FP-LMTO & 19.6 & 13.6 & 22.0 & 9.56 \\
& PAW [Ne]$3s^2$ & 20.4 & 14.0 & 22.9 & 9.61 \\ 
& PAW [Ar]+corr & 20.2 & 13.7 & 22.9 & 9.42 \\ 
& & & & & \\
7.55 & FP-LMTO & 13.4 & 10.1 & 15.9 & 7.19 \\
& PAW [Ne] & 14.2 & 10.5 & 16.4 & 7.16 \\
& PAW [Ar]+corr & 14.0 & 10.3 & 16.4 & 7.15 \\
& & & & & \\  
9.70 & FP-LMTO & 6.33 & 6.37 & 8.93 & 4.89 \\ 
& PAW [Ne]$3s^2$ & 7.23 & 6.43 & 9.52 & 4.65 \\
& PAW [Ar]+corr & 7.35 & 6.49 & 9.64 & 4.74 \\
\end{tabular}
\caption{Zone boundary phonon frequencies (THz units) of f.c.c. Fe from the 
present PAW calculations compared with the FP-LMTO results of 
Ref.~\protect\cite{soderlind96}. Results are shown for three values of the
volume per atom $V_0$. PAW results are given for both the Ne$3s^2$ core and
the Ar core plus effective pair-potential correction (see text).}
\label{tab:fccphon}
\end{table} 

\begin{table}
\begin{tabular}{llcccc}
                               &          &  67   & 127   & 241  & 499 \\
\tableline
$D (10^{-9}$ m$^2$~s$^{-1}$) & PAW [Ne]$3s^2$ & $ 4.8 \pm 0.2$ & & &  \\ 
                               & PAW [Ar] + corr & $5.2 \pm 0.2$  & & & \\ 
                               & Ref. model  & 6.1 & 6.2 & 6.4 & 6.5 \\
& & & & & \\
$\eta$ (mPa~s)                 & PAW [Ne]$3s^2$ &  $5.5\pm 2 $ &  & &  \\
                               & PAW [Ar] + corr & $8.5\pm 1$ & &  & \\
                               & Ref. model & 7.0 & 7.5 & 7.0 & 6.5 \\
\end{tabular}
\caption{Comparison of {\em ab-initio} and
reference-model values of the diffusion
coefficient $D$ and the viscosity $\eta$ for
simulated {\em l}-Fe systems containing
67, 127, 241 and 499 atoms at $T = 4300$~K and $\rho = 10700$~kg~m$^{-3}$.
PAW results are given for the Ne$3s^2$ core and
the Ar core with pair-potential correction (see text).}\label{tab:finitesize}
\end{table}

\begin{table}
\begin{tabular}{l|ccccc}
 &  \multicolumn{5}{c}{$\rho$ (kg m$^{-3}$) } \\
\tableline
 $T$ (K) & 9540 & 10700 & 11010  & 12130 & 13300 \\
\tableline
  3000 &   60   &       &     &     &     \\  
       &        &       &     &     &     \\  
  4300 &        & 132   &     &     &     \\  
       &        & (135) &     &     &     \\  
  5000 &        & 140   &     &     &     \\  
       &        & (145) &     &     &     \\  
  6000 &  90    &  151  & 170 & 251 & 360 \\  
       &        & (155) &(170) & (240) & (335) \\  
  7000 &        &  161  & 181 & 264 & 375 \\  
       &        &       &     & (250) & (350) \\  
  8000 &        &  172  & 191 & 275 & 390 \\  
       &        &       &     &  & (360) \\  
\end{tabular}
\caption{Pressure (GPa units) calculated in the
full set of 16 {\em ab-initio}
simulations of {\em l}-Fe. Experimental values (in parenthesis) are from 
Ref.~\protect\cite{anderson94}. }\label{tab:pressure}
\end{table}

\begin{table}
\begin{tabular}{ll|ccccc}
 & & \multicolumn{5}{c}{$\rho$~(kg~m$^{-3}$) } \\
\tableline
     & $T$~(K) & 9540 & 10700   & 11010  & 12130 & 13300 \\
\tableline
$D$ ($10^{-9}$ m$^2$~s$^{-1}$) & 3000 &  $4.0\pm 0.4$ &   &    &     &    \\  
           & 4300 &        & $5.2 \pm 0.2$ & &     &     \\  
           & 5000 &        & $7.0\pm 0.7$ &  &  &     \\  
& 6000 &  $14 \pm 1.4$ & $10\pm 1$ & $9 \pm 0.9$ & $6\pm 0.6$ & $5\pm 0.5$\\  
& 7000 &  & $13 \pm 1.3$   & $11 \pm 1.1$  & $9 \pm 0.9$  & $6 \pm 0.6 $\\  
& & & & & \\
$\eta$ (mPa~s)    & 3000 &   $6\pm 3$  &     &     &    \\  
                  & 4300 &        & $8.5\pm 1$ &     &      \\  
                  & 5000 &        & $6 \pm 3 $ &     &      \\  
        & 6000 &  $2.5\pm 2$ & $5 \pm 2 $ & $7\pm 3$ & $8\pm 3$ & $15\pm 5$\\ 
                 & 7000 &  & $4.5\pm 2$ &  $4\pm 2$ & $8\pm 3$ & $10\pm 3$ \\ 
\end{tabular}
\caption{The diffusion
coefficient $D$ and the viscosity $\eta$ from ab-initio
simulations of {\em l}-Fe
at a range of temperatures and densities. The error estimates come from 
statistical uncertainty due to the short duration of the simulation.}
\label{tab:d_eta}
\end{table}

\clearpage

\protect\twocolumn[
\hsize\textwidth\columnwidth\hsize\csname@twocolumnfalse\endcsname
\centerline{\psfig{figure=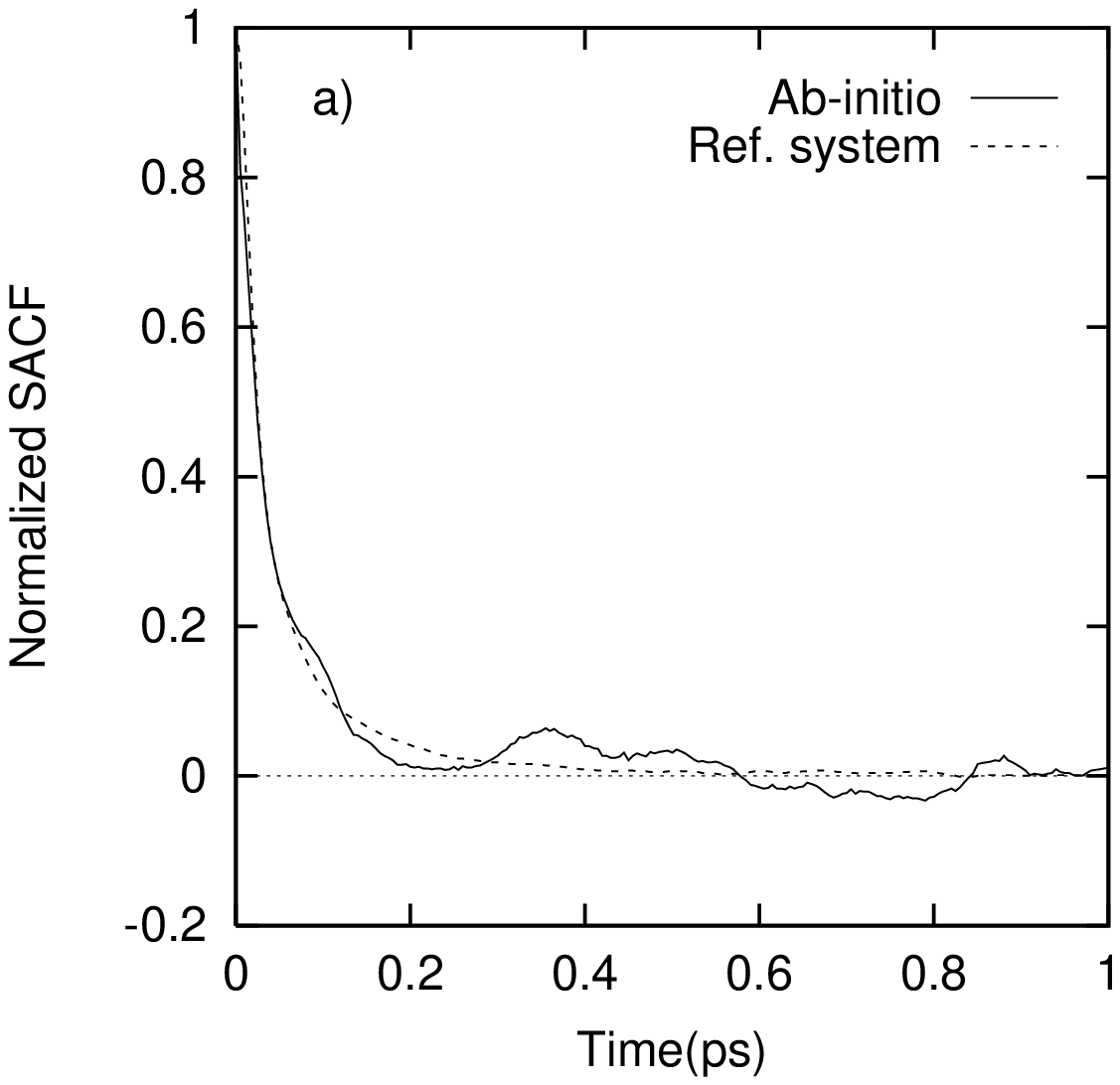,height=3.0in}
\psfig{figure=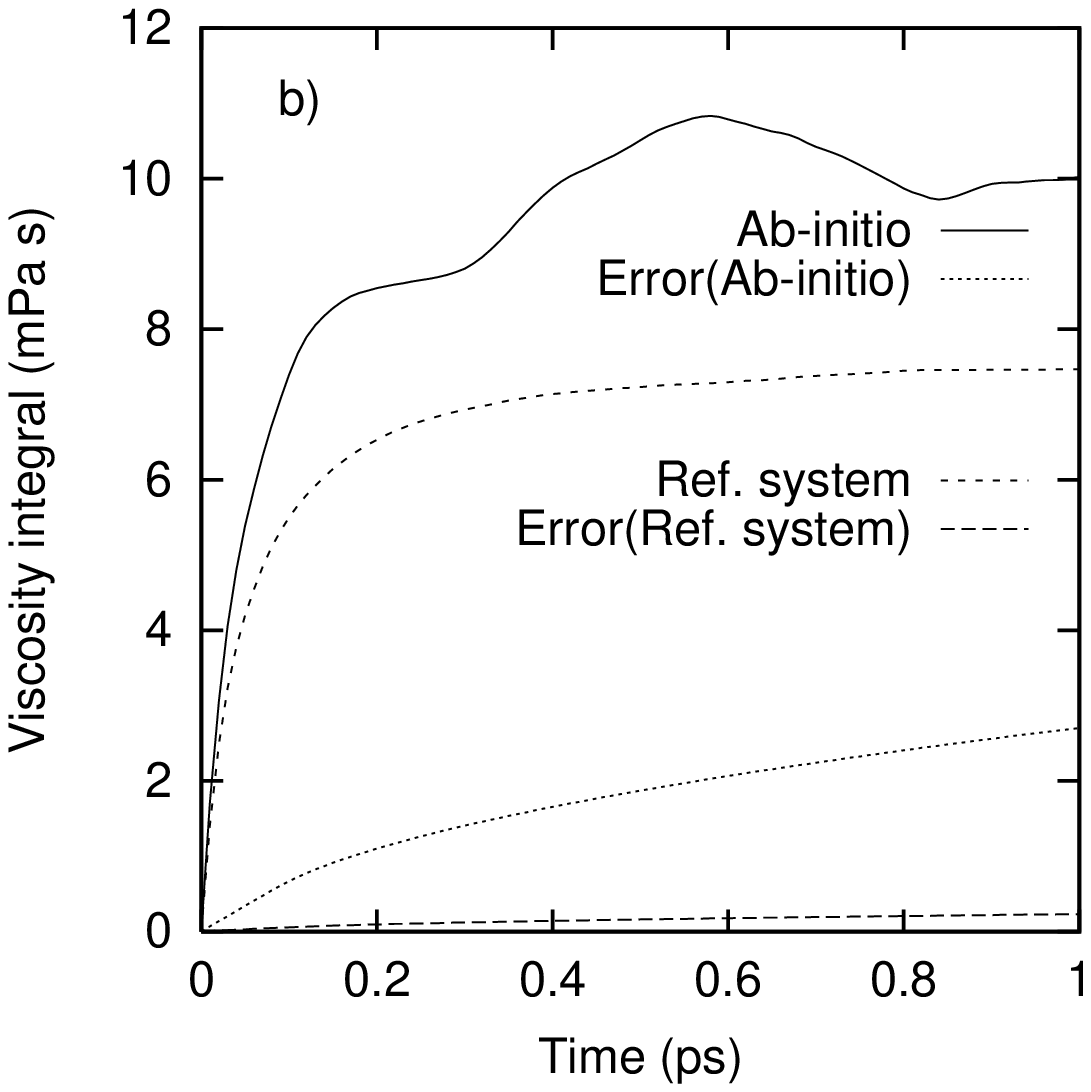,height=3.0in}}
\begin{figure}
\caption{(a) Stress autocorrelation function (see
eqn. (\ref{eqn:sacf})) normalised to its $t = 0$ value of liquid Fe at
$T = 4300$~K and $\rho = 10700$~kg~m$^{-3}$. Solid and short-dashed
curves show {\em ab-initio} and reference-system results. (b)
Viscosity integral (see text) of {\em l}-Fe at the same thermodynamic
state. Solid and short-dashed curves as in panel (a); dotted and
long-dashed curves show statistical errors. }\label{fig:sacf}
\end{figure}
]

\protect\twocolumn[
\hsize\textwidth\columnwidth\hsize\csname@twocolumnfalse\endcsname
\centerline{\psfig{figure=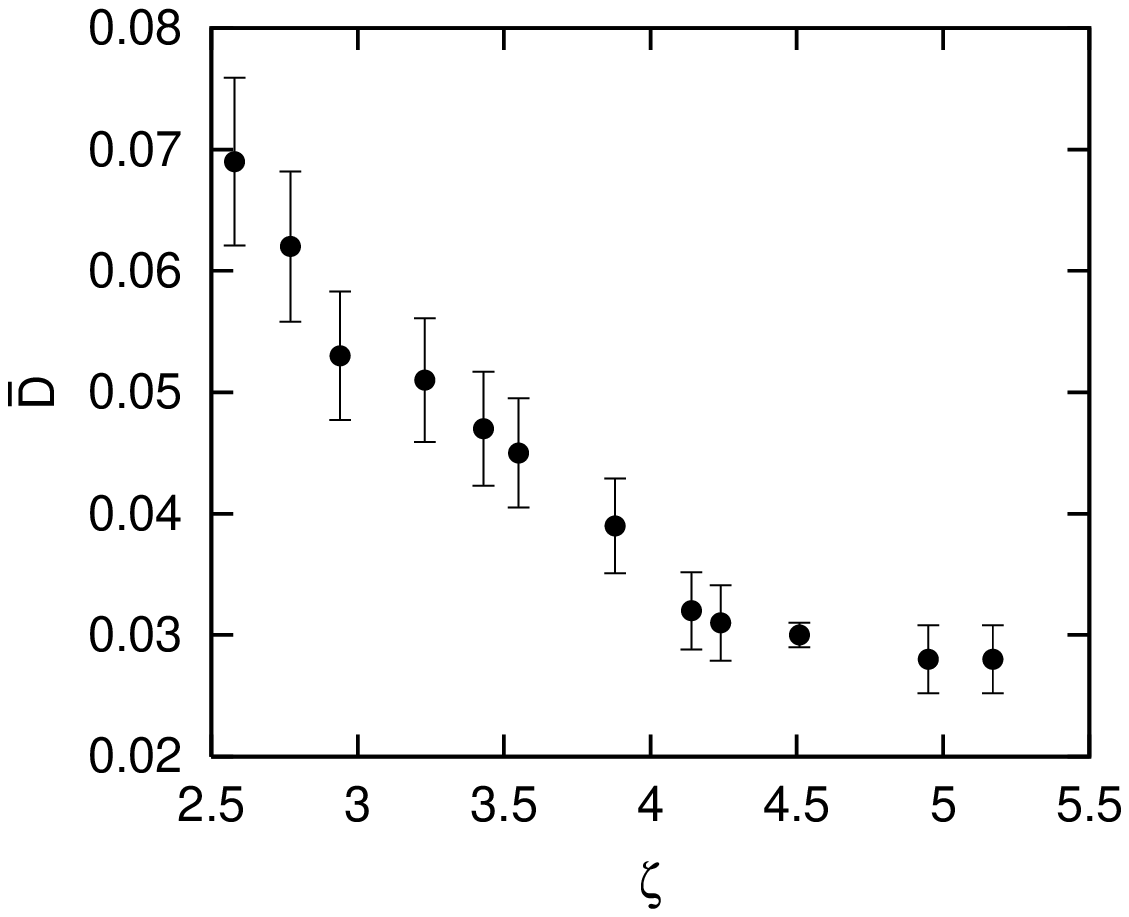,height=2.5in}\psfig{figure=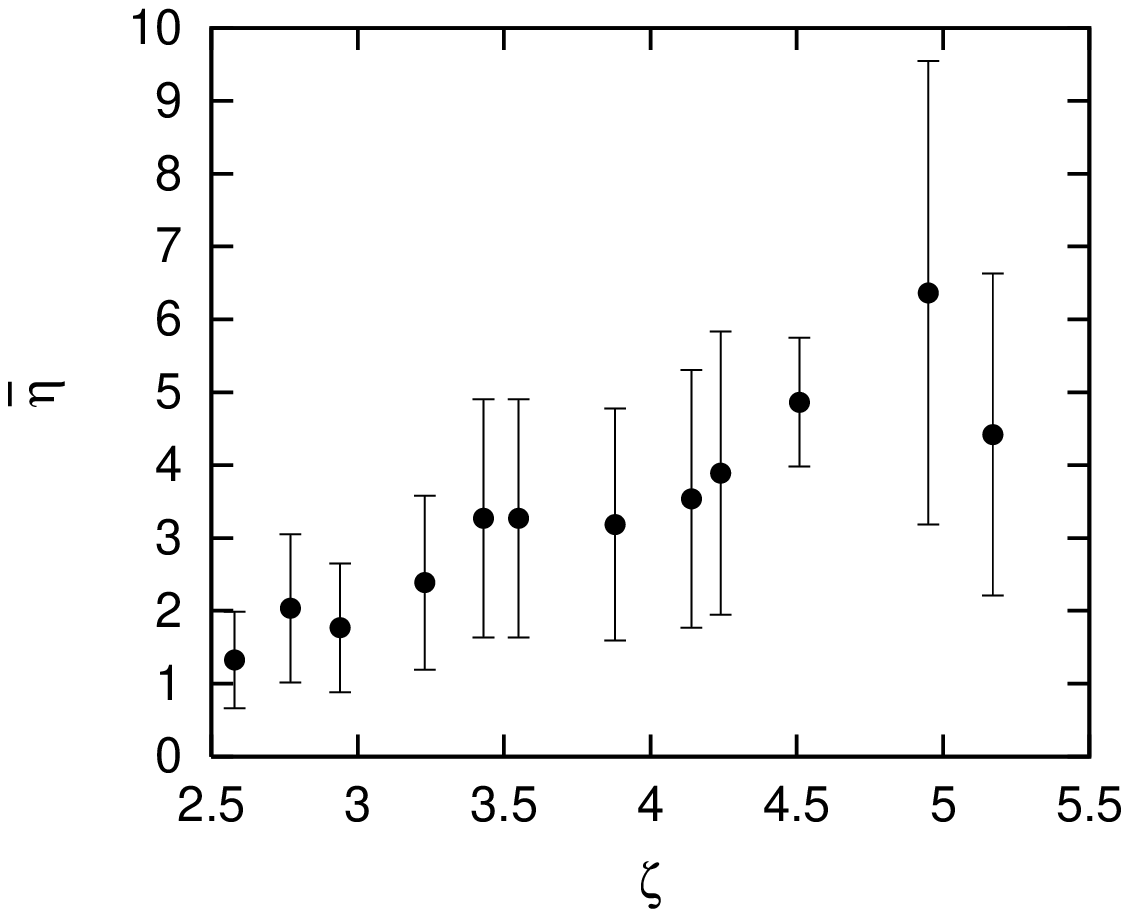,height=2.5in}}
\begin{figure}
\caption{Reduced diffusion coefficient $\bar{D}$ and viscosity 
$\bar{\eta}$ as a function of reduced state variable $\zeta$ (see text)
from ab-initio simulations of {\em l}-Fe at 12
thermodynamic states. Error bars
show statistical uncertainty on each point.}\label{fig:d_eta}
\end{figure}
]

\end{document}